\documentclass[aps,prl,twocolumn,amsmath,amssymb, superscriptaddress,tightenlines]{revtex4}
\usepackage{graphicx}
\usepackage{epsfig}
\usepackage{dcolumn}
\usepackage{bm}
\usepackage{amssymb}
\usepackage{bm}
\usepackage{amsmath, bm}
\usepackage{upgreek}
\usepackage[colorlinks,citecolor=blue,linkcolor=blue,hyperindex]{hyperref}

\begin{document}

\title{Realization of Supersymmetry and Its Spontaneous Breaking in Quantum Hall edges}
\author{Ken K. W. Ma, Ruojun Wang, and Kun Yang}
\affiliation{National High Magnetic Field Laboratory and Department of Physics, Florida State University, Tallahassee, Florida 32306}
\date{\today}


\begin{abstract}
Supersymmetry (SUSY) relating bosons and fermions plays an important role in unifying different fundamental interactions in particle physics. Since no superpartners of elementary particles have been observed, SUSY, if present, must be broken at low-energy. This makes it important to understand how SUSY is realized and broken, and study their consequences. We show that an $\mathcal{N}=(1,0)$ SUSY, arguably the simplest type, can be realized at the edge of the Moore-Read quantum Hall state. Depending on the absence or presence of edge reconstruction, both SUSY-preserving and SUSY broken phases can be realized in the same system, allowing for their unified description. The significance of the gapless fermionic Goldstino mode in the SUSY broken phase is discussed.
\end{abstract}

\maketitle


\textbf{Introduction} - Supersymmetry (SUSY)~\cite{Golfand1971} is a theoretical construction that relates bosons and fermions. Its presence may resolve the hierarchy problem between weak interaction and gravity~\cite{SUSY-primer}, and bring in new particles as candidates of dark matter~\cite{Fayet1976, Fayet1977}. Furthermore, the superstring theory being one of the leading candidates to unify all four fundamental interactions relies on SUSY~\cite{Polchinski}. All these make the idea of SUSY very appealing. However, no superpartners of standard model elementary particles have been observed yet. Therefore, SUSY must be broken at low-energy, even if it is an underlying symmetry of nature.

In addition to looking for possible signatures of SUSY from high-energy experiment (mostly at the Large Hadron Collider)~\cite{Peskin, LHC}, proposals of realizing SUSY and a related concept, supersymmetric quantum mechanics~\cite{SQM-book}, in other systems have been proliferating recently, ranging from quantum optics~\cite{Andreev-QO, Tomka-QO, Hirokawa-QO}, cold atoms~\cite{Snoek2005, Snoek2006, Yu-Yang2008, Yu-Yang2010, Lai-Yang2015, Blzaiot2015, Bradlyn-Gromov, Blzaiot2017}, lattice models~\cite{Nicolai1976, Nicolai1997, Fendley-JPA2003, Fendley-PRL2003, Lee2007, Sachdev2010, Bauer2013, Hsieh2016, Sannomiya2016, Sannomiya2019,  Nayga2019}, and condensed matters~\cite{Foda, SUSY-5/2-edge, Qi-TSC, ZKM-TSC, ZK-TSC, WZ-graphene, Vishwanath2014, Ponte2014, Berg2015, Affleck2015, Yao2015, Yao-SQED2017, Yao-edge2017, SUSY-FQHE, SUSY-SYK, Juven-SPT1, Juven-SPT2}. In many of these cases SUSY emerge at isolated critical points. It would be highly desirable to explore systems that exhibit SUSY in stable phases and, in particular, can be driven through a phase transition at which SUSY breaks spontaneously.

In this Letter, we demonstrate the existence of an $\mathcal{N}=(1,0)$ SUSY in the edge theory of the Moore-Read (MR) state~\cite{MR1991}, when the bosonic and fermionic edge modes, which are superpartners of each other, have the same velocity. The MR state is a candidate state exhibiting non-Abelian topological order in quantum Hall (QH) states at the filling factor $\nu=1$ for bosons~\cite{Gunn2000, Cooper2001, Regnault2003, Jain2005, Regnault2007, Viefers2008}, or half-integer filling factors for fermions~\cite{Willett1987, Pan1999, Eisenstein-bilayer, Suen-bilayer, Eisenstein-7/2, Liu-7/2, Falson2015, Falson2018, bi-graphene-Ki, bi-graphene-Kim, bi-graphene-Li, bi-graphene-Zibrov, mono-graphene-Zibrov, high-LL-graphene, WSe2-half}. We show that the unreconstructed MR edge has a supersymmetric ground state. We further demonstrate that SUSY is broken by edge reconstruction, which can be triggered by softening of the edge confinement potential. This allows for a unified study of SUSY and its breaking in the same system. The spontaneous breaking of SUSY results in a massless Goldstino fermion mode in the reconstructed phase, which in turn is the indication of underlying SUSY. 

\textbf{Chiral supersymmetry in Moore-Read edge} - We start by reviewing the edge theory of the MR state described by the action~\cite{MR-edge}, 
\begin{eqnarray} \label{eq:free-edge}
S_0
=\int dtdx~\left[-\partial_x\varphi(\partial_t+v_b\partial_x)\varphi+i\psi(\partial_t+v_f\partial_x)\psi\right],
\end{eqnarray}
in terms of a chiral Bose mode $\varphi$ and a copropagating Majorana fermion mode $\psi$. The rescaled Bose mode $\varphi=\phi/\sqrt{4\pi\nu}$ is defined, so $\rho(x)=\partial_x\varphi=\sqrt{\pi/\nu}(\partial_x\phi/2\pi)$, where $\partial_x\phi/2\pi$ is the particle number density along the edge, parametrized by coordinate $x$.

We define the SUSY transformation:
\begin{align}
\label{eq:SUSY-bose}
\delta\varphi
&=\epsilon\psi,
\\
\label{eq:SUSY-MF}
\delta\psi
&=-i\epsilon\partial_x\varphi,
\end{align}
where $\epsilon$ is an infinitesimal Grassmannian parameter. When $v_b=v_f=v$, $S_0$ is invariant under the SUSY transformation up to a boundary term,
\begin{eqnarray}
\delta S_0
=-\epsilon\int dtdx~ \frac{\partial}{\partial x}
\left[\psi\left(\partial_t+v\partial_x\right)\varphi\right].
\end{eqnarray}
In a QH liquid, $\partial_x\varphi$ and $\partial_t\varphi$ satisfy the periodic boundary conditions (PBCs), $\partial_x\varphi(x)=\partial_x\varphi(x+L)$ and
$\partial_t\varphi(x)=\partial_t\varphi(x+L)$. Thus, $\delta S_0=0$ if the fermionic mode satisfies the PBC, $\psi(x)=\psi(x+L)$ also. Here, $L$ is the length of the edge, or the circumference in a disk-shaped sample. The boundary condition of $\psi(x)$ depends on the number of non-Abelian quasiparticles in the bulk of the QH liquid. If this number is odd (even), then $\psi(x)$ has a periodic (antiperiodic) boundary condition~\cite{MR-edge, Fisher-BC}. Even in the case of anti-PBC, the symmetry-breaking boundary term has a negligible effect for sufficiently large $L$. This condition can be reached in a real sample. Hence, we assume the boundary term vanishes throughout our work. Meanwhile, we do not include quasihole or quasiparticle in our discussion, so that additional low-energy excitations can be avoided. It is natural for them to be absent in the ground state of a quantum Hall liquid.

To better understand SUSY of Eq.~\eqref{eq:free-edge}, we consider the corresponding Hamiltonian:
\begin{eqnarray} \label{eq:free-edge-H}
H_0=\int \left[v_b(\partial_x\varphi)^2-iv_f\psi\partial_x\psi\right]dx.
\end{eqnarray}
Now, $\varphi(x)$ and $\psi(x)$ become field operators that satisfy the quantization rules
\begin{align}
\left[\varphi(x), \partial_y\varphi(y)\right]
&=-\frac{i}{2}\delta(x-y),
\\
\left\{\psi(x),\psi(y)\right\}
&=\frac{1}{2}\delta(x-y).
\end{align}
The SUSY transformation in Eqs.~\eqref{eq:SUSY-bose} and~\eqref{eq:SUSY-MF} is generated by the supercharge,
\begin{eqnarray} \label{eq:supercharge}
Q=2\int dx~\psi(x)\partial_x\varphi(x)
=2\sum_{k>0}\left(\rho_k^\dagger\psi_k+ \rho_k\psi_k^\dagger\right),
\end{eqnarray}
which is Hermitian and satisfies $Q=Q^\dagger$. To switch to the momentum-space representation, we define the Fourier transform of the Bose field,
\begin{eqnarray}
\varphi(x)=\frac{1}{\sqrt{L}}\sum_k e^{ikx}\varphi_k,
\end{eqnarray}
and a similar transform for the fermion field. Note
$\rho_k=ik\varphi_k$ being the Fourier transform of $\rho(x) = \partial_x \varphi$ satisfies the Kac-Moody algebra
$\left[\rho_k, \rho_p^\dagger\right]=k\delta_{k,p}/2$. By writing $H_0$ in the momentum-space representation:
\begin{eqnarray} \label{eq:free-edge-k}
H_0
=2\sum_{k>0}\left[v_b\rho_k^\dagger\rho_k+ v_fk\psi_k^\dagger \psi_k\right]
+\frac{v_b-v_f}{2}\sum_{k>0}k,
\end{eqnarray}
it is easy to show $[H_0, Q] = 0$ when $v_b = v_f$, which is SUSY by definition. In particular,
one finds that the ground state of $H_0$ is the vacuum of chiral bosons and Majorana fermions. We denote this vacuum state as $|\text{vac}\rangle$. It satisfies
$\rho_k|\text{vac}\rangle=0$ and $\psi_k|\text{vac}\rangle=0$ for all $k>0$, thus the ground state is annihilated by $Q$:
\begin{eqnarray}
Q|\text{vac}\rangle=0
\label{eq:SUSY ground state}
\end{eqnarray}
and therefore supersymmetric. It is also clear the bosonic mode $\varphi$ and Majorana fermion mode $\psi$ are superpartners of each other, and when $v_b=v_f$, they have identical spectra, which is a consequence and direct manifestation of SUSY.

The SUSY along the MR edge discussed above is an $\mathcal{N}=(1,0)$ SUSY~\cite{Hull-Witten, Gates}, which is the simplest type consisting of only one real supercharge. It is also known as an $\mathcal{N}=1/2$ SUSY~\cite{Sakamoto}. In string theory, it is common to define the light-cone coordinates $x^{\pm}=(x^0\pm x^1)/\sqrt{2}$ to describe left and right movers. The $\mathcal{N}=(1,0)$ SUSY algebra is generated by a supercharge $Q_+$ satisfying $\left\{Q_+, Q_+\right\}=2P_+=2i\partial/\partial x^+$. This is possible only if the Lorentz group has an irreducible representation, in which left-moving and right-moving fermions can be decoupled. One possible case is the $(1+1)$ dimensions. In this case, chiral Majorana fermions (usually known as Majorana-Weyl fermions in high energy physics) exist~\cite{footnote}. A more general $\mathcal{N}=(p,q)$ SUSY can be constructed if one also includes the supercharge $Q_{-}$ satisfying $\left\{Q_-, Q_-\right\}=2P_{-}$. A general discussion on this can be found in the work by Hull and Witten~\cite{Hull-Witten}.

To relate our work with SUSY in a more transparent manner, as well as to prepare for introduction of more general SUSY Lagrangians that allow for edge reconstruction, we summarize the superfield formalism for the $\mathcal{N}=(1,0)$ SUSY in the $(1+1)$ dimensions below~\cite{Hull-Witten, Gates, Sakamoto}. The superfield lives in a superspace
$\mathbb{R}^{2|1}=(t,x,\theta^+)$, with the fermionic coordinate
$\theta^+$ satisfying
\begin{eqnarray}
(\theta^+)^2=0
~,~
\frac{\partial\theta^+}{\partial\theta^+} =1
~,~
\frac{\partial x}{\partial\theta^+} = 0
~,~
\int d\theta^+ =\frac{\partial}{\partial\theta^+}.
\end{eqnarray}
The Bose field $\varphi(x)$ and Majorana fermion field $\psi(x)$ in the previous discussion can be grouped into a scalar superfield,
\begin{eqnarray}
\Phi=\varphi(x)+\theta^+\psi(x).
\end{eqnarray}
In the present work, we define the superderivative as
\begin{eqnarray}
\mathcal{D}
=\frac{\partial}{\partial\theta^+}-i\theta^+\frac{\partial}{\partial x}.
\end{eqnarray}
and the supercharge acting in the superspace as
\begin{eqnarray}
\mathcal{Q}
=-i\frac{\partial}{\partial\theta^+}+\theta^+\frac{\partial}{\partial x}.
\end{eqnarray}
This satisfies $\left\{\mathcal{Q},\mathcal{D}\right\}=0$ and $\left[\mathcal{Q}, \partial_x\right]=0$. In addition, $\mathcal{D}$ and $\mathcal{Q}$ satisfy
$\left\{\mathcal{D}, \mathcal{D}\right\}=\left\{\mathcal{Q}, \mathcal{Q}\right\}=-2i\partial_x =2\mathcal{P}$~\cite{footnote2}, where $\mathcal{P}$ is the total momentum operator acting in the superspace. The SUSY transformation in Eqs.~\eqref{eq:SUSY-bose} and~\eqref{eq:SUSY-MF} can be reformulated as $\delta\Phi=i\epsilon\mathcal{Q}\Phi$. Using the superfield formalism, one can construct SUSY-invariant actions systematically:
\begin{eqnarray} \label{eq:S-SUSY}
S=\int dtdxd\theta^+ ~ W(\Phi, \mathcal{D}^m\Phi, \partial_t^n\Phi).
\end{eqnarray}
Here, $W$ can be an arbitrary function of the superfield and its derivatives. The orders $m$ and $n$ need not be equal, and the spatial derivative of $\Phi$ is actually included since
$\mathcal{D}^2\Phi=-i\partial_x\Phi$. The function $W$ can be expanded in a power series of $\theta^+$. The term with $\theta^+$ changes by a spatial derivative under the SUSY transformation~\cite{SUSY-primer}. Hence, $\delta S$ is a boundary term. If it vanishes, then $S$ is invariant. We assume $v_b=v_f=v$ in the following discussion. The edge theory~\eqref{eq:free-edge} can be rewritten as
\begin{eqnarray}
S_0
=-i\int dtdxd\theta^+~ \mathcal{D}\Phi(\partial_t+v\partial_x)\Phi.
\end{eqnarray}
This verifies that $S_0$ is invariant under the SUSY transformation generated by
$\mathcal{Q}$.

It is important to clarify the differences between our work and other common SUSY models in high energy physics. First, the SUSY that appears in the present work is not an extension of the Lorentz symmetry. It is reflected in the definitions of $\mathcal{D}$, $\mathcal{Q}$, and the corresponding SUSY algebra. All of them just contain $\partial_x$ but not $\partial_t$. Therefore, space and time are not treated equally in our case. We emphasize that this feature originates from the definition of SUSY algebra itself, but not the seeming absence of Lorentz invariance due to the chirality of our theory. In fact, a recent work pointed out that the Floreanini-Jackiw action of chiral boson can be a gauge-fixed version of a manifestly Lorentz invariant theory~\cite{Townsend2020}. Second, the Bose field $\varphi$ in our present case is chiral, which differs from the nonchiral Klein-Gordon field in high energy physics. Since all fields in our system are chiral, it is fair to call the SUSY in the present case as a chiral $\mathcal{N}=(1,0)$ SUSY. Note that the same model was mentioned briefly by Sonnenschein~\cite{Sonnenschein1988}. Finally, the SUSY in the present work is realized in the (1+1)-D spacetime. In superstring theory, the spacetime needs to be ten dimensional~\cite{Polchinski}. The $\mathcal{N}=(1,0)$ SUSY in 2D is a symmetry in the worldsheet or a compactified manifold~\cite{Hull-Witten, Gates}, which cannot be observed directly.

In 2D electron gas systems like that hosted by GaAs, the MR state is a candidate to describe the edge of the $\nu=5/2$ fractional QH state~\cite{Willett1987, Pan1999}. In such systems numerical work showed that $v_b\gg v_f$ due to the long-range Coulomb interaction and the charged nature of the Bose mode $\varphi$, and the fact that $\psi$ is a neutral mode~\cite{Wan2006, Wan2008}. Hence, it is unlikely to realize the above SUSY there. On the other hand QHE can also be realized in charge-neutral ultracold atomic systems~\cite{Gunn2000, Cooper2001, Regnault2003, Jain2005, Regnault2007, Viefers2008}. In particular, previous numerical works have shown robust formation of the MR state at $\nu=1$ for bosons with short-range repulsive interaction~\cite{Gunn2000, Cooper2001, Regnault2003, Jain2005, Regnault2007}. In this case the condition $v_b=v_f$ can be realized by tuning the confinement (or trapping) potential. We would like to emphasize however, even if $v_b\ne v_f$, which means the theory of Eq.~\eqref{eq:free-edge} does not have SUSY, its ground state, which is the vacuum of the bosonic and fermionic modes and {\em independent} of the velocities, {\em remains} supersymmetric and satisfies Eq. (\ref{eq:SUSY ground state}). We thus find the ground state is {\em more} (super)symmetric than the Hamiltonian if $v_b\ne v_f$, and we can have a ground state that possesses SUSY {\em without} fine-tuning, as long as edge reconstruction does not occur. This further demonstrates the relevance and importance of $\mathcal{N}=(1,0)$ SUSY. We now turn to the case of edge reconstruction. \\

\textbf{Spontaneous SUSY breaking in edge reconstruction} -  The above discussion showed that the simplest version of the MR edge possesses SUSY. Can SUSY be broken in the same system? The answer is yes, and as we show below, it is closely tied with another very important and common piece of physics, namely edge reconstruction~\cite{CSG-edge, Chamon-edge,Wan-Yang-Rezayi, Wan-Rezayi-Yang}, which we now review.

In addition to having the above simple edge structure, it is possible for a QH liquid to undergo edge reconstructions. This may originate from the interplay of Coulomb interaction and the confining potential in electronic systems~\cite{CSG-edge, Chamon-edge, Wan-Yang-Rezayi, Wan-Rezayi-Yang}, and similar physics is relevant to cold atom systems. To illustrate the physical idea, we digress a bit and review the field theoretical description of edge reconstruction in the Laughlin state. When edge reconstruction is absent, the Laughlin state has a single chiral Bose mode $\varphi$ along the edge~\cite{Wen-book}. The corresponding low-energy excitations along the edge are chiral bosons with momenta $k\gtrapprox 0$. By taking momentum dependence of electron-electron interaction into account, Yang showed that the (bosonic) edge excitation has a nonlinear energy dispersion~\cite{Kun2003}:
\begin{align}
\epsilon(k)=v(k-ak^3+bk^5 +\cdots)~,~ k>0.
\end{align}
Here, all coefficients $v$, $a$, and $b$ are positive. When $a>a_c=2\sqrt{b}$, $\epsilon(k)$ attains its global minimum at a nonzero momentum $k_0$. In this situation, the many-body ground state of the system is no longer given by the vacuum with no chiral bosons. Instead, these bosons can occupy quantum states with momenta $k\approx k_0$ to minimize the energy of the system. This leads to an increase in the momentum along the edge, and corresponds to an edge reconstruction. To have a stable ground state, it is necessary to include a repulsive interaction between the bosons, which originates from nonlinearity of the edge confinement potential. Taking all the above into consideration, the Laughlin edge is described by the Hamiltonian
\begin{eqnarray} \label{eq:eff-Laughlin}
H_{\rm eff}
=2\sum_{k>0} v(1-ak^2+bk^4)\rho_k^\dagger \rho_k
+\tilde{V}.
\end{eqnarray}
The symbol $\tilde{V}$ denotes the momentum-space representation of the repulsive interaction~\cite{Kun2003}
\begin{eqnarray} \label{eq:int-Bose}
V=\int \left[u_3(\partial_x\varphi)^3+u_4(\partial_x\varphi)^4\right]~dx.
\end{eqnarray}

Is it possible to construct a supersymmetric field theory for edge reconstruction in the MR state? The answer is positive and can be achieved from the superfield formalism. We observe that a part of the following superfield action,
\begin{align}
\nonumber
&S_{\rm int}
\\ \nonumber
=&-i\int dtdxd\theta^+~
\left[u_3(\partial_x\Phi)^2+u_4(\partial_x\Phi)^3\right]\mathcal{D}\Phi
\\ \nonumber
=&-\int dt~V+i\int dtdx~\left[2u_3\partial_x\varphi+3u_4(\partial_x\varphi)^2\right]
\psi\partial_x\psi
\\ \nonumber
=&-\int dt~V -\int dt~V_{\rm BF}
\\
=&-\int dt~V_{\rm int},
\end{align}
reproduces the action of $V$ for the Bose field in Eq.~\eqref{eq:int-Bose}. Moreover, the result shows that an additional term $V_{\rm BF}$ is required, such that the action
$S_{\rm int}$ remains invariant under the SUSY transformation in Eqs.~\eqref{eq:SUSY-bose} and~\eqref{eq:SUSY-MF}. Based on the above discussion, we propose a more general supersymmetric theory for the MR edge:
\begin{align}
H_{\rm MR}
=2v\sum_{k>0} (1-ak^2+bk^4)(\rho_k^\dagger \rho_k + k\psi_k^\dagger\psi_k)
+\tilde{V}_{\rm int}.
\end{align}
Here, $\tilde{V}_{\rm int}$ is the momentum-space representation of $V_{\rm int}$. Thanks to this term, the supersymmetric edge theory is an interacting field theory, which cannot be diagonalized analytically as in Eq.~\eqref{eq:free-edge-k}. It is interesting that the theory requires an interaction between the Bose field and the Majorana fermion field. The assumption of having the same dispersion for bosonic and fermionic modes may require a delicate fine-tuning. Nevertheless, it may be still achievable since both confining potential and interaction in cold atom systems can be tuned with a high degree of flexibility.

Following the logic in the Laughlin state, the MR state undergoes an edge reconstruction when $a>a_c$. We denote the resultant reconstructed MR edge ground state as $|\Phi_0\rangle$, which contains a finite number of bosons and fermions with momenta $k>0$ and unlike the vacuum state $|\text{vac}\rangle$, has a finite momentum density~\cite{Kun2003}. Now, we show that the SUSY is broken spontaneously in $|\Phi_0\rangle$. A generic state $|\Phi\rangle$ satisfies
\begin{eqnarray} \label{eq:Q-Q}
\langle\Phi| \left\{Q, Q\right\} |\Phi\rangle
\geq 0.
\end{eqnarray}
Since $\left\{Q,Q\right\}=2H_0/v =2P$, $|\text{vac}\rangle$ is the {\em only} zero momentum state that satisfies $\langle\text{vac}| \left\{Q, Q\right\} |\text{vac}\rangle
= 0$ and therefore supersymmetric. Instead
$\langle\Phi_0| \left\{Q,Q\right\}|\Phi_0\rangle > 0$ and from Eq.~\eqref{eq:Q-Q},
$Q|\Phi_0\rangle \neq 0$. Because $|\Phi_0\rangle$ is not annihilated by the supercharge $Q$, this ground sate breaks the SUSY spontaneously. \\

\textbf{Existence of Goldstino mode} - Since SUSY is a fermionic symmetry, its spontaneous breaking leads to a gapless Goldstone fermion mode, also known as Goldstino~\cite{Fayet-Iliopoulos1974, Fayet-Goldstino}. The zero-momentum Goldstino state is defined as $Q|\Phi_0\rangle$~\cite{Weinberg-3}. This state has the same energy as $|\Phi_0\rangle$ because $[H_{\rm MR},Q]=0$, and is therefore a Majorana zero mode. We note Majorana zero modes are of tremendous interest recently, especially in the context of topological phases and topological quantum computation~\cite{MZM-TQC}. In our case the situation is very different; it is due to symmetry breaking, but {\em not} of the Landau type, because SUSY is an unusual type of symmetry whose breaking is {\em not} described by Landau theory.

One qualitative difference between $Q|\Phi_0\rangle$ and the Majorana zero modes in topological phases is the former is a member of a dispersing (and nonchiral) Goldstino mode,
whose wave function is
\begin{align}
\nonumber
R_q^\dagger |\Phi_0\rangle
&=2\int dx~e^{-iqx} \psi\partial_x\varphi |\Phi_0\rangle
\\
&=2\sum_{p>0}\left(\rho_p\psi_{p-q}^\dagger +\rho_p^\dagger \psi_{p+q}\right)
|\Phi_0\rangle,
\end{align}
where $q$ is the momentum of the mode. Different from the case of a Bose-Fermi mixture in which the bosons form a BEC~\cite{Yu-Yang2008}, the process of turning a fermion into a boson or vice versa cannot be implemented easily in the present case. At the same time, we should clarify that this inability does not imply the Goldstino fermion is undetectable. In fact any single fermion process should couple to the the Goldstino mode. 

We now use the variational principle to deduce the energy dispersion of the Goldstino:
\begin{eqnarray}
\Delta(q)
=\frac{\langle\Phi_0 | R_q (H_{\rm MR}-E_0) R_q^\dagger |\Phi_0\rangle}
{\langle\Phi_0 | R_q R_q^\dagger |\Phi_0\rangle}
\sim \alpha q^2,
\end{eqnarray}
where $\alpha>0$. To arrive at the $q^2$ dispersion, we have expanded $R_q$ and
$R_q^\dagger$ in the power series of $q$. Since $[H_{\rm MR}-E_0, Q]=0$ and
$(H_{\rm MR}-E_0)|\Phi_0\rangle=0$, the nonvanishing term has a leading order of $q^2$. The existence of such a gapless, quadratic fermionic mode is (in principle) a clear indication of underlying SUSY, even though it is spontaneously broken. Its presence can (for example) be detected through its contribution to low-temperature ($T$) specific heat: $c_{\rm G}\sim \sqrt{T/\alpha}$, which {\em dominates} the linear $T$ contributions from edge modes with linear $q$ dispersions (the generic situation, with or without edge reconstruction~\cite{Kun2003}). In experiment, the system always has a temperature much lower than the one associated with the energy gap in the bulk. This gap can only be overcome when the system size is exponentially large. For systems with finite and not-too-large sizes, the thermal transport is dominated by the edge contribution as demonstrated in experiment. Thus, the contribution from bulk excitations is negligible~\cite{finalnote}. \\

\textbf{Concluions and outlook} - To summarize, we demonstrated that the $\mathcal{N}=(1,0)$ SUSY can be realized at the edge of the Moore-Read quantum Hall state. This system can support both supersymmetric and SUSY broken phases, with the transition between them triggered by edge reconstruction. Underlying SUSY is manifested by the existence of a gapless Goldstino fermion with a $q^2$ dispersion when it is broken spontaneously.

Despite its theoretical simplicity, the unreconstructed MR edge has not been realized experimentally. We hope our observation that it has a SUSY ground state will provide additional motivation for experimentalists to look for systems that can realize it. If successful, it provides a “proof of principle” that SUSY can be realized in simple systems made of either just bosons (at $\nu=1$), or just fermions (at $\nu=1/2$). We believe such “proof of principle” can be of great conceptual value, as it demonstrates SUSY can emerge (reasonably) naturally in simple systems, with no or minimal fine-tuning. To the best of our knowledge, no other proposals apart from string theory have been introduced to realize the $\mathcal{N}=(1,0)$ SUSY. Implementing our proposal in condensed matter or cold atom systems can open the door to study this simplest kind of SUSY experimentally.

Finally, we note that the {\em bulk} excitations in the Moore-Read state, namely, the Girvin-MacDonald-Platzman mode and the neutral fermion mode can be viewed as superpartners~\cite{Gromov}. It would be very interesting to pursue their unified description via a supersymmetric generalization of the bimetric theory~\cite{Gromov-bimetric}. Also, it will be tempting to examine the possibility of realizing {\em local} supersymmetry (namely, supergravity) and the associated massive gravitino~\cite{Fayet-gravitino1, Fayet-gravitino2} in QH systems, given the recent excitement of gravitational analogies~\cite{finalrefs}. \\

K. Y. thanks Andrey Gromov for a useful conversation. The authors also acknowledge Pierre Fayet, Juven Wang, and Jan Behrends for useful comments. This research was supported by the National Science Foundation Grant No. DMR-1932796, and performed at the National High Magnetic Field Laboratory, which is supported by the National Science Foundation Cooperative Agreement No. DMR-1644779, and the State of Florida.


\begin{thebibliography}{99}
\bibitem{footnote}
In this paper, chirality is defined as the traveling direction of Majorana fermions. This definition may differ from the general definition in high energy physics.
\bibitem{footnote2}
We still regard this as an $\mathcal{N}=(1,0)$ SUSY. The difference in a prefactor of $-\sqrt{2}$ originates from the definition of light-cone coordinates. In this paper, we follow the usual convention in quantum mechanics with momentum operator $P=-i\partial/\partial x$.
\end{thebibliography}
\end{document}